\documentclass[11pt]{article}
\usepackage{pstricks}
\usepackage{pst-vue3d}
\usepackage{pst-3dplot}
\usepackage{apacite}
\usepackage{amsfonts}
\usepackage{amsmath}
\usepackage{amsthm}
\newcommand{\ket}[1]{|#1\rangle}
\newcommand{\braket}[1]{|#1\rangle \langle #1|}
\newcommand{\sa}[1]{#1 _{\mathrm{sa}}}
\newcommand{\bhs}{\mathcal{B}(\mathcal{H})_{\mathrm{sa}}}

\newcommand{\nbit}{\mathcal{B}(\mathbb{C}^{n})}

\newcommand{\card}[1]{|#1|}
\newcommand{\tp}[2]{|\langle #1|#2\rangle |^{2}}
\newcommand{\ratio}[3]{\mathbf{p}_{#1}(#2/#3)}

\newcommand{\twoqubit}{\mathcal{B}(\mathbb{C}^{2}\otimes \mathbb{C}^{2})}
\newcommand{\alg}[1]{\mathfrak{#1}}
\newcommand{\hil}[1]{\mathcal{#1}}
\newcommand{\bh}{\mathcal{B}(\mathcal{H})}
\newcommand{\norm}[1]{\|#1\|}

\newcommand{\hsnorm}[1]{\|#1\| _{2}}
\newcommand{\abs}[1]{|#1|}
\newcommand{\tr}[1]{\mathrm{Tr}(#1)}
\newcommand{\co}[1]{\mathrm{co}(#1)}
\newcommand{\extreme}[1]{\partial _{e}#1}
\newcommand{\ext}[1]{\partial _{e}#1}
\newcommand{\face}[2]{\mathrm{face(}#1,#2\mathrm{)}}
\newtheorem{lemma}{Lemma}

\newtheorem*{thm*}{Theorem}
\newtheorem*{rt}{Root Theorem}

\newtheorem{thm}{Theorem}

\theoremstyle{definition}
\newtheorem*{defn}{Definition}
\theoremstyle{remark}

\title{A note on information theoretic characterizations of physical theories}
\author{Hans Halvorson\thanks{{\tt hhalvors@princeton.edu.} \ This is version 2.} \\ {\small Department of Philosophy,
  Princeton University} }
\date{ }
\begin{document}
\maketitle
\begin{abstract}
  Clifton, Bub, and Halvorson [{\it Foundations of Physics} 33,
  1561--1591, (2003)] have recently argued that quantum theory is
  characterized by its satisfaction of three information-theoretic
  axioms.  However, it is not difficult to construct apparent
  counterexamples to the CBH characterization theorem.  In this paper,
  we discuss the limits of the characterization theorem, and we
  provide some technical tools for checking whether a theory
  (specified in terms of the convex structure of its state space)
  falls within these limits.
\end{abstract}

\section{Introduction}
Some would like to argue that quantum information theory has
revolutionary implications for the philosophical foundations of QM
(see, e.g., \citeNP{bub,fuchs}).  Whether or not this claim is true,
there is no doubt that quantum information theory presents us with new
perspectives from which we can approach traditional questions about
the interpretation of QM.  One such question asks whether there are
natural physical postulates that capture the essence of QM ---
postulates that tell us what sets QM apart from other physical
theories, and in particular from its predecessor theories.  The advent
of quantum information theory suggests that we look for
\emph{information-theoretic} postulates that characterize (i.e., are
equivalent to) QM.

A positive answer to this question has been supplied by Clifton, Bub,
and Halvorson \citeyear{bch}.  Clifton, Bub and Halvorson (CBH) show
that, within the $C^{*}$-algebraic framework for physical theories,
quantum theories are singled out by their satisfaction of three
information-theoretic axioms: 1. no superluminal information transfer
via measurement; 2. no broadcasting;\footnote{For the case of pure
  states, broadcasting reduces to cloning.}  and 3.  no
unconditionally secure bit commitment.  Nonetheless, the creative
thinker will have little trouble concocting a ``theory'' that
satisfies these three axioms, but which does not entail QM (see
\citeNP{spekkens,smolin}).  Such toy theories might be thought to show
that the three information-theoretic axioms are not sufficient to
recover the full structure of QM.

Since the CBH characterization theorem is a valid mathematical result,
there is a problem of application here --- these apparent
counterexamples must not satisfy the premises of the theorem.  Besides
the three information-theoretic axioms, the only other premise of the
theorem is the $C^{*}$ assumption (i.e., the assumption that a
theory's observables be representable by the self-adjoint operators in
a $C^{*}$-algebra).  However, in specific cases, it may be difficult
to ascertain whether or not a theory satisfies the $C^{*}$ assumption.
In particular, since the axioms for $C^{*}$-algebras are rather
intricate, and some of these axioms have no direct physical
interpretation (e.g., the $C^{*}$-algebraic product of non-commuting
observables does not correspond to any physical operation on
observables), there is a prima facie difficulty in relating the
$C^{*}$ assumption to specific features of a physical theory.

In this paper, we address the difficulty in determining whether a toy
theory satisfies the $C^{*}$ assumption.  In particular, it is
sometimes easier to ascertain the convex structure of the state space
of a theory (i.e., which states are mixtures of which other states)
than to ascertain the algebraic structure of the observables of that
theory.  Furthermore, due to the deep mathematical results of Alfsen
et al. (detailed in \citeNP{geometry}), specifying the convex
structure of the state space of a theory is sufficient to determine
whether that theory can be formulated within the Jordan-Banach (JB)
algebraic framework.  Since a theory permits a $C^{*}$-algebraic
formulation only if it permits a JB algebraic formulation, showing
that a theory does not permit a JB algebraic formulation is sufficient
to show that it falls outside of the range of validity of the CBH
theorem.

The structure of this paper is as follows.  In Section~2, we review
the basics of the theory of JB algebras, and of the dual (but more
general) theory of convex sets.  We also prove a ``Root Theorem,''
which forms the basis for our results in subsequent sections.  In
Section 3, we address the worry that the three information-theoretic
axioms are not sufficient to entail QM.  In particular, we look at a
certain class of toy theories that satisfies the axioms, and we show
that these toy theories do not permit a JB algebraic formulation.  In
Section~4, we consider a class of theories that are locally
quantum-mechanical, but which, unlike QM, do not have nonlocally
entangled states.  We show that the simplest of these theories does
not permit a JB algebraic formulation; and we adduce considerations
which indicate that no such theory permits a JB algebraic formulation.

\section{The JB algebraic framework for physical theories}

\subsection{Jordan-Banach algebras}

The CBH theorem shows that among the theories within the
$C^{*}$-algebraic framework, quantum theories are precisely those that
satisfy the three information-theoretic axioms.  One limitation of
this result is that it excludes from consideration those theories that
employ real or quaternionic Hilbert spaces (and so the result does not
shed any light on the physical significance of the choice of the
underlying field for a Hilbert space).  We can get past this
limitation by moving to the broader JB algebraic framework.

Let $M_{n}(\mathbb{K})$ be the set of $n\times n$ matrices over
$\mathbb{K}$, where $\mathbb{K}=\mathbb{R},\mathbb{C}$, or
$\mathbb{H}$ (the quaternions).  The set $H_{n}(\mathbb{K})$ of
Hermitian matrices in $M_{n}(\mathbb{K})$ is a vector space over
$\mathbb{R}$.  If we set $A\circ B=\frac{1}{2}(AB+BA)$, where $AB$ is
the usual matrix product of $A$ and $B$, then it follows that
\begin{equation} ((A\circ A)\circ B)\circ A = (A\circ A)\circ (B\circ
  A) .\label{jordan} \end{equation}
The matrix algebra
$H_{n}(\mathbb{K})$ with product $\circ$ is the
prototype for the notion of a Jordan algebra: a Jordan algebra is any
real vector space equipped with a commutative (not necessarily
associative), bilinear product $\circ$ satisfying Eqn.~\ref{jordan}.    

For an element $A\in H_{n}(\mathbb{C})$, we define 
\begin{equation} \norm{A}=\sup \{ \norm{Ax}:x\in \mathbb{C}^{n},\;
  \norm{x}=1\}, \end{equation}
where the norm on the right is the vector norm on $\mathbb{C}^{n}$.  
This norm is complete in the sense that for any
Cauchy sequence $\{ A_{i} \}$ in $H_{n}(\mathbb{C})$, there is an
$A\in H_{n}(\mathbb{C})$ such that $\lim _{i}\norm{A_{i}-A}=0$.  That
is, $H_{n}(\mathbb{C})$ is a Banach space.  Furthermore, the norm
satisfies the inequalities:
\begin{equation} \norm{A\circ B}\leq \norm{A}\norm{B},\quad
  \norm{A\circ A}=\norm{A}^{2},\quad \norm{A\circ A}\leq
  \norm{(A\circ A)+(B\circ B)},\label{banach} \end{equation}
for all $A,B$ in $H_{n}(\mathbb{C})$.  In general, a Jordan-Banach
(JB) algebra is a Jordan algebra that is complete relative to some
norm satisfying Eqns.~\ref{banach}.  
  
States on $H_{n}(\mathbb{C})$ are given, in the first place, by
(equivalence classes of) unit vectors in $\mathbb{C}^{n}$.  In
particular, if $\ket{\alpha}$ is a unit vector in $\mathbb{C}^{n}$,
then $\langle \alpha |A|\alpha \rangle$ gives the expectation value of
$A$ in the state $\ket{\alpha}$.  Note that the map $A\mapsto \langle
\alpha |A|\alpha \rangle$ on $H_{n}(\mathbb{C})$ is linear and
continuous.  Furthermore, $\langle \alpha |I|\alpha \rangle =1$, where
$I$ is the identity matrix, and and $\langle \alpha |(A\circ A)|\alpha
\rangle \geq 0$ for any $A$.  Generally, we define a state of a JB
algebra $\alg{A}$ to be a linear and continuous mapping $\omega
:\alg{A}\rightarrow \mathbb{R}$ such that $\omega (I)=1$ and $\omega
(A\circ A)\geq 0$ for all $A$ in $\alg{A}$.

The state space $K$ of a JB algebra $\alg{A}$ is a convex set; that
is, if $x$ and $y$ are states, and $\lambda \in (0,1)$, then $\lambda
x+(1-\lambda)y$ defines a state in a natural way.  The set $K$ also
carries two standard topologies.  First, a net $\{ \omega _{a} \}$ of
states converges in the weak* topology to a state $\omega$ just in
case the numbers $\{ \omega _{a}(A):A\in \alg{A} \}$ converge
pointwise to the numbers $\{ \omega (A):A\in \alg{A} \}$.  Since $K$
is a weak* closed subset of the unit ball of the Banach space dual
$\alg{A}^{*}$ (all continuous linear functionals on $\alg{A}$), the
Alaoglu-Bourbaki theorem \cite[Thm.~1.6.5]{kr} entails that $K$ is
weak* compact.  Second, $K$ inherits the standard norm topology from
$\alg{A}^{*}$.  A net $\{ \omega _{a} \}$ converges in norm to
$\omega$ just in case the numbers $\{ \omega _{a}(A) :A\in \alg{A} \}$
converge uniformly to the numbers $\{ \omega (A):A\in \alg{A} \}$.
Thus, the norm topology on $K$ is always finer that the weak*
topology.  In the finite dimensional case, pointwise convergence
entails uniform convergence, and so the weak* and norm topologies are
equivalent.  But in the infinite dimensional case, $K$ will not
typically be compact in the norm topology.  (For example, the state
space of the JB algebra $\sa{\bh}$ of all self-adjoint operators on an
infinite dimensional Hilbert space $\hil{H}$ is not compact in the
norm topology.)

There is a canonical mapping from the category of $C^{*}$-algebras
into the category of JB algebras.  Indeed, if $\alg{A}$ is a
$C^{*}$-algebra, and $\sa{\alg{A}}$ is the real vector space of
self-adjoint operators in $\alg{A}$, then $\sa{\alg{A}}$ with the
symmetric product is a JB algebra (\citeNP[Thm.~1.1.9]{land}).
Furthermore, the state space of $\alg{A}$ is affinely isomorphic (see
the definition below) to the state space of $\sa{\alg{A}}$.  In
contrast, the nonassociative JB algebra $H_{2}(\mathbb{R})$ has linear
dimension $3$, whereas there is no $C^{*}$-algebra $\alg{A}$ such that
$\sa{\alg{A}}$ is a $3$-dimensional, nonassociative JB algebra.
($\sa{\alg{A}}$ is associative iff $\alg{A}$ is abelian.)  Therefore,
$H_{2}(\mathbb{R})$ is not isomorphic to the self-adjoint part of
$C^{*}$-algebra, and the JB algebraic framework is genuinely broader
than the $C^{*}$-algebraic framework.

\subsection{Convex sets}

All JB algebra state spaces are convex sets.  But the converse is not
true --- not all convex sets are JB algebra state spaces.  We now
briefly recall some of the main definitions in the theory of convex
sets.

A point $x$ in a convex set $K$ is {\it extreme} just in case for any
$y,z\in K$ and $\lambda \in (0,1)$, if $x=\lambda y+(1-\lambda )z$,
then $x=y=z$.  We let $\extreme{K}$ denote the set of extreme points
in $K$. If $K$ is the state space of an algebra, we also call extreme
points {\it pure states}.  A subset $F$ of a convex set $K$ is said to
be a {\it face} just in case $F$ is convex, and for any $x\in F$, if
$x=\lambda y+(1-\lambda )z$ with $\lambda \in (0,1)$, then $y\in F$.
Clearly the intersection of an arbitrary family of faces is again a
face. For $x,y\in K$, we let $\face{x}{y}$ denote the intersection of
all faces containing $\{ x,y \}$.  A pair of faces $F,G$ in $K$ is
said to be {\it split} if every point in $K$ can be expressed uniquely
as a convex combination of points in $F$ and $G$.  A convex set $K$ is
said to be a {\it simplex} if mixed states have unique decompositions
into pure states.  More precisely, $K$ is a simplex if for all
$w,x,y,z\in \ext{K}$, when
\begin{equation} \lambda w+(1-\lambda )x= \mu y+(1-\mu
  )z ,\end{equation} with $\lambda ,\mu \in (0,1)$, then either $w=y$ or
$w=z$.  (This definition differs slightly from the standard definition; see \cite[p.~8]{states}.)  
If $K$ and $L$ are convex sets, a mapping $\phi
:K\rightarrow L$ is an {\it affine isomorphism} just in
case $\phi$ is bijective, and 
\begin{equation} \phi (\lambda x+(1-\lambda)y)=\lambda \phi (x)+(1-\lambda)\phi
(y) , \end{equation} for all $x,y\in K$ and $\lambda \in (0,1)$.  If there is an affine
isomorphism $\phi$ from $K$ onto $L$, then $K$ and $L$ are said to be
{\it affinely isomorphic}. 

\subsection{The root theorem}

Drawing on the results of Alfsen et al., we now derive some easily
checked necessary conditions for a theory to admit a JB algebraic
formulation.  (In this theorem and subsequently, we let $B^{n}$ denote
the closed unit ball in $\mathbb{R}^{n}$.)
 
\begin{rt} Let $K$ be a convex set.  If $K$ is affinely
  isomorphic to the state space of a JB algebra, then:
\begin{enumerate}
\item For any distinct $x,y\in \ext{K}$, $\face{x}{y}=B^{n}$ for some
  $n\geq 1$.
\item If $K$ is not a simplex, then for any distinct $x,y\in \ext{K}$,
  $\face{x}{y}=B^{n}$ for some $n\geq 2$.
\item If $x,y\in \ext{K}$ are connected by a norm-continuous path,
  then $\face{x}{y}=B^{n}$ for some $n\geq 2$. \label{path}
\end{enumerate}
\end{rt}

The statement of (3) could use some clarification: since we have not
made any assumptions about a topology on $K$, saying that $\ext{K}$ is
connected does not really make sense.  However, if $K$ is affinely
isomorphic to the state space $K'$ of a JB algebra, then there is a
map $\phi :K\rightarrow K'$.  Thus, (3) should be understood as
referring to the topology on $K$ that is induced, via the mapping
$\phi$, by the norm topology on $K'$.

\begin{proof} (1.) The first statement is a non-trivial result (Corollary 5.56 in
  \citeNP{geometry}) that depends on a number of lemmas.  Due to space
  constraints, we just sketch the structure of the proof for the
  simple case where $K$ is a subset of a finite-dimensional vector
  space.
  
  Suppose that $K$ is the state space of a JB algebra $\alg{A}$.
  Since $x,y$ are pure states, they correspond to minimal projection
  operators $P,Q\in \alg{A}$, and $\face{x}{y}$ is the state space of
  the ``projected'' algebra
\begin{equation} \alg{A}_{P\vee Q}=\{ (P\vee Q)A(P\vee Q):A\in \alg{A} \}
  .\end{equation}  The identity of
$\alg{A}_{P\vee Q}$ (namely, $P\vee Q$) is the sum of two
  orthogonal projections $P$ and $R=(P\vee
  Q)-P$.  We now consider the two cases where this projected
algebra is associative or nonassociative.  If $\alg{A}_{P\vee Q}$ is associative, then it is
isomorphic to the algebra of real valued functions on a two-point set,
and its state space consists of two pure states and their convex
combinations.  That is, the state space of $\alg{A}_{P\vee Q}$, and
therefore $\face{x}{y}$, is isomorphic to $B^{1}$.  If
$\alg{A}_{P\vee Q}$ is nonassociative, then in fact the center of
$\alg{A}_{P\vee Q}$ is trivial.  By the comparison theorem for
projections, there is a symmetry $U\in
\alg{A}_{P\vee Q}$ (that is, $U\circ U=I$) such that \begin{equation}
2(U\circ (P\circ U))-P = Q.
\end{equation}
Thus, the identity in $\alg{A}_{P\vee Q}$ is the sum of two
``exchangeable'' minimal projections.  Finite dimensional JB algebras
with this property have been completely classified (see
\citeNP[Prop.~3.37]{geometry}), and their state spaces are isomorphic
to $B^{n}$, for some $n\geq 2$.
  
  (2.) If $K$ is not a simplex, then there are $w,x,y,z\in \ext{K}$
  such that
\begin{equation} \lambda w+(1-\lambda )x=\mu y+(1-\mu )z ,\end{equation}
where $\lambda ,\mu \in (0,1)$, $w\neq y$ and $w\neq z$.  But then $w$
is an extreme point in $\face{y}{z}$.  We know from part 1 that
$\face{y}{z}=B^{n}$, for some $n\geq 1$.  Since there are three
distinct extreme points of $\face{y}{z}$, it follows that $n\geq 2$.

(3.) We prove the contrapositive.  Suppose that $x,y\in \ext{K}$, and
$\face{x}{y}=B^{1}$.  Then there are split faces $F,G$ of $K$ such
that $x\in F$ and $y\in G$ \cite[Lemma 5.54]{geometry}.  Let $U=F\cap
\ext{K}$ and let $V=G\cap \ext{K}$.  Since $F$ and $G$ are closed in
the norm topology \cite[Prop.~1.29]{states}, $U$ and $V$ are closed in
$\ext{K}$.  Since $\ext{K}\subseteq F\cup G$, it follows that
$\ext{K}=U\cup V$, and $U$ and $V$ are open in $\ext{K}$.  Since $x\in
U$ and $y\in V$, there is no continuous path in $\ext{K}$ connecting
$x$ and $y$.
\end{proof}

\section{Sufficiency of the axioms}

The state space of a quantum system has {\it ambiguous mixtures} ---
i.e., mixed states with more than one decomposition into pure states
--- and this fact is responsible for some of the interesting
information-theoretic features of QM.  For example, the BB84
\cite{bb84} bit commitment protocol is perfectly concealing because
two distinct ensembles can be absolutely identical relative to a local
observer (since these ensembles correspond to the same quantum state).
Thus, in order to find a toy theory that simulates some of the
information-theoretic features of QM, it would be natural to look for
simple theories with ambiguous mixtures.

One such theory has been recently described by \citeA{spekkens}.
(\citeA{smolin} proposes a different sort of theory that satisfies the
three axioms.  We look at Smolin's theory in \cite{hb}.)  The state
space $S$ of Spekkens' theory (for a local system) has exactly seven
points: the pure states correspond to the unit vectors $\{
e_{i},-e_{i}:i=1,2,3 \}$ in $\mathbb{R}^{3}$, and the mixed state
corresponds to the origin $\mathbf{0}$ in $\mathbb{R}^{3}$.  In order
to equip $S$ with partial binary operations corresponding to
superposition and mixture, we identify $S$ with a subset of the Bloch
sphere.  That is, $\bf{0}$ is an equal mixture of $e_{i}$ and
$-e_{i}$, for $i=1,2,3$.  However, $S$ does not contain unequal
mixtures of $e_{i}$ and $-e_{i}$, nor does $S$ contain mixtures of
$e_{i}$ and $e_{j}$ when $i\neq j$.  Similarly, $e_{i}$ and $-e_{i}$
can be superposed with equal weights to obtain any of the states in
$\{ e_{j},-e_{j}:j\neq i \}$.  However, $e_{i}$ and $-e_{i}$ cannot be
superposed with unequal weights, nor can $e_{i}$ be superposed with
$e_{j}$ when $i\neq j$.

Since $S$ is not convex, it is obviously not the state space of a JB
algebra.  However, the failure of convexity can be easily remedied by
passing to the modified theory that allows arbitrary mixtures of
Spekkens' states --- i.e., the theory whose state space is the convex
hull $K=\co{S}$ of $S$.\footnote{Presumably, Spekkens would want to
  say that the ``transition probability'' between $e_{1}$ and $e_{2}$
  is $\frac{1}{2}$, since $e_{2}$ is supposed to be an equally
  weighted superposition of the orthogonal states $e_{1}$ and
  $-e_{1}$.  However, the natural geometric transition probability of
  $e_{1}$ and $e_{2}$, relative to the convex set $K$, is $0$.  (See
  the definition of affine ratio below.)  In particular, if every
  affine function from $K$ into $\mathbb{R}$ corresponds to an
  observable (as is usually assumed in the convex sets approach), then
  there is a measurement that can distinguish with certainty between
  $e_{1}$ and $e_{2}$; for example, the measurement corresponding to
  the function $f(x)=\frac{1}{2}(1+(e_{1}-e_{2}+e_{3})\cdot x)$.
  Presumably, then, Spekkens would wish to impose some restriction on
  the set of observables.}  Clearly, $K$ has ambiguous mixtures, and
has exactly six pure states.  We now show that convex sets of this
sort are not state spaces of JB algebras.

\begin{thm} Let $K$ be a convex set, and suppose that $K$ is affinely
  isomorphic to the state space of a JB algebra.  If $K$ is not a
  simplex, then $\card{\ext{K}}\geq \card{\mathbb{R}}$. \label{finite}
\end{thm}

\begin{proof}  Suppose that $K$ is not a simplex.  Then part 1 of the
  Root Theorem entails that there are $x,y\in \ext{K}$ such that
  $\face{x}{y}=B^{n}$, with $n\geq 2$.  Since every extreme point in
  $\face{x}{y}$ is an extreme point in $K$, we have
  $\card{\ext{K}}\geq \card{\ext{B^{n}}}=\card{\mathbb{R}}$.
\end{proof}

Spekkens' theory does not permit a JB algebraic formulation, and {\it
  a fortiori}, does not not permit a $C^{*}$-algebraic formulation.
So, this theory falls outside the range of validity of the CBH
theorem.  Nonetheless, since Spekkens' theory has no obvious physical
pathologies, it would be a interesting test case for the claim that
physical theories should permit, at the very least, a JB algebraic
formulation.  

\section{Independence of the axioms}

All parties agree that, in the presence of the $C^{*}$ assumption
(i.e., the assumption that theories permit a $C^{*}$-algebraic
formulation), the three information-theoretic axioms entail QM.  It
seems, then, that the real question is whether the $C^{*}$ assumption
is true, warranted, reasonable, or something like that.
Unfortunately, it seems that it would be extremely difficult to give a
decisive answer to this question.

However, there is reason to think that the $C^{*}$ assumption is doing
too much work in the CBH theorem.  In particular, given the $C^{*}$
assumption, QM is a logical consequence of the first two axioms alone.
In fact, given the $C^{*}$ assumption, the no bit commitment axiom is
a logical consequence of the no superluminal signaling and no cloning
axioms.  We prove this fact here with the one simplifying assumption
that the relevant algebras are actually von Neumann algebras (i.e.,
the algebras act on some concrete Hilbert space $\hil{H}$, and are
closed in the weak-operator topology).

\begin{thm} Suppose that $\alg{A}$ and $\alg{B}$ are von Neumann algebras.  If
  the composite system $(\alg{A},\alg{B})$ satisfies the no
  superluminal signaling and no cloning axioms, then:
\begin{enumerate}
\item $(\alg{A},\alg{B})$ has nonlocally entangled states; and
\item $(\alg{A},\alg{B})$ satisfies the no bit commitment axiom.
\end{enumerate}
  \label{independence}
\end{thm}

\begin{proof} 
  Suppose that the pair $(\alg{A},\alg{B})$ satisfies the no
  superluminal signaling and no cloning axioms.  On the one hand, the
  no superluminal signaling axiom entails that observables in
  $\alg{A}$ commute with observables in $\alg{B}$ \cite[Thm.~1]{bch}.
  On the other hand, the no cloning axiom entails that $\alg{A}$ and
  $\alg{B}$ are nonabelian \cite[Thm.~2]{bch}.
  
  (1.) When $\alg{A}$ and $\alg{B}$ are nonabelian, a theorem by
  Landau \citeyear{landau} shows that there are nonlocally entangled
  (indeed, Bell correlated) states across $(\alg{A},\alg{B})$.
  
  (2.) By the generalized HJW theorem \cite{halvorson}, for any two
  equivalent measures $\mu ,\nu$ on the state space of $\alg{B}$
  (i.e., these measures correspond to the same quantum state), there
  is an entangled state $\psi$ of $(\alg{A},\alg{B})$ such that either
  $\mu$ or $\nu$ can be prepared from $\psi$ by local operations on
  $\alg{A}$.  So, for any bit commitment protocol for
  $(\alg{A},\alg{B})$, if the protocol is concealing, then it is not
  binding. \end{proof}

\subsection{The Schr*dinger theory}

Theorem~\ref{independence} is somewhat surprising.  From an apparently
local axiom (no cloning), it follows that there are nonlocally
entangled states.  In slogan form: {\it any locally quantum mechanical
  theory is nonlocal}.  We may contrast this result with
Schr{\"o}dinger's claim that there should be a locally quantum
mechanical theory without entangled states.  Schr{\"o}dinger says:

\begin{quote}
  Indubitably, the situation described here [in which there are
  nonlocally entangled states] is, in present QM, a
  necessary and indispensable feature.  The question arises, whether
  it is so in Nature too.  I am not satisfied about there being
  sufficient experimental evidence for that\dots
  
  It seems worth noticing that the paradox could be avoided by a very
  simple assumption, namely if the situation after [two systems]
  separating were described by the expansion
\begin{equation} c_{1}|01\rangle +c_{2}|10\rangle , \label{phase} \end{equation} 
but with the additional statement that the knowledge of the
\emph{phase relations} between the complex constants $c_{1}$ and
$c_{2}$ has been entirely lost in consequence of the process of
separation.  This would mean that not only the parts, but the whole
system, would be in the situation of a mixture, not of a pure state.
\dots it would utterly eliminate the experimenter's influence on the
state of that system which he does not touch.
  
This is a very incomplete description and I would not stand for its
adequateness.  But I would call it a possible one, until I am told,
either why it is devoid of meaning or with which experiments it
disagrees.

(\citeNP[pp.~451--452]{schro}.  Eqn.~\ref{phase} has been adapted to
the present discussion.)
\end{quote}

When Schr{\"o}dinger speaks of the state in Eqn.~\ref{phase}, but with
``the knowledge of the phase relations'' lost, he presumably means the
mixed state
\begin{equation} \abs{c_{1}}^{2} |01\rangle \langle
  01|+\abs{c_{2}}^{2}|10\rangle \langle 10|  .\end{equation}
Thus, Schr{\"o}dinger's hope is that the true theory will
turn out to be locally quantum mechanical, but with some sort of
selection rule that prohibits superposition of product states for
systems that are spacelike separated.

We now know --- due to experimental verification of the violation of
Bell's inequality --- that Schr{\"o}dinger's hoped-for theory
disagrees with experiment.  But, Theorem~\ref{independence} shows that
Schr{\"o}dinger's hoped-for theory is ``devoid of meaning'' --- well,
at least if all meaningful theories admit a $C^{*}$-algebraic
formulation.  But can Schr{\"o}dinger's hope be realized within the
broader JB algebraic framework?  While we do not currently know the
answer to this question, we will proceed to show that the answer is
negative in one particularly simple case.

Consider the simplest composite quantum system, a pair of qubits.  Of
course, we cannot simply throw away the nonlocally entangled states
without doing violence to the linear structure of
$\mathbb{C}^{2}\otimes \mathbb{C}^{2}$.  But since the complement of
the set of nonlocally entangled states is a convex set, we can throw
away the entangled states and still end up with a theory with a convex
state space.  More precisely, recall that a density operator $D\in
\twoqubit$ is a {\it pure product state} just in case $D=E\otimes F$,
where $E,F$ are projections onto rays in $\mathbb{C}^{2}$.  The set of
pure product states is a closed subset of the pure state space of
$\twoqubit$; as such, it is a closed, bounded subset of
$\mathbb{R}^{15}$.  (The set of $4\times 4$ Hermitian complex matrices
has real dimension 16, and the subset of positive, trace-1 matrices
has real dimension 15.)  Let $K$ denote the set of convex combinations
of pure product states; in other words, $K$ is the space of
\emph{separable states} of $\twoqubit$, and corresponds to a compact
convex subset of $\mathbb{R}^{15}$.  (For some results on the geometry
of $K$, see \cite{thirring}.)

Since $K$ is a convex set, it gives a genuine theory in the convex
sets approach; we call this theory the {\it Schr*dinger theory}.
(This ad hoc construction is for conceptual purposes only; we do not
think that Schr{\"o}dinger really had this theory in mind when he
expressed his hope for an alternative to QM.)  The
observables of the Schr*dinger theory are the elements of the real
vector space $A(K)$ of affine functions from $K$ to $\mathbb{R}$.  The
expectation value of observable $f\in A(K)$ in state $x\in K$ is
$f(x)$.  Clearly, each self-adjoint operator $A\in \twoqubit$ gives an
observable for the Schr*dinger theory via the mapping $x\mapsto
\tr{xA}$.

In order to clarify the information-theoretic properties of the
Schr*dinger theory, we need to define a notion of transition
probability for arbitrary convex sets (see \citeNP{mielnik};
\citeNP[Prop.~2.8.1]{land}).

\begin{defn} Let $K$ be a convex set, and let $A(K)$ be the
  set of affine functions from $K$ into $\mathbb{R}$.  If $x,y\in
  \ext{K}$, then the affine ratio (or transition probability) of $x$
  and $y$ relative to $K$ is given by
\[ \ratio{K}{x}{y}\: =_{\mathrm{def}} \: \inf \{
f(y); f \in A(K),\; \mathrm{range}(f)\subseteq [0,1],\; \text{and} \;
f(x)=1 \, \} .\] \end{defn}

The affine ratio has a natural geometrical interpretation.  In
particular, if $K$ is a contained in the vector space $V$, then each
affine function $f:V\rightarrow \mathbb{R}$ foliates $V$ into a family
of hyperplanes $\{ f^{-1}(t)\} _{t\in \mathbb{R}}$.  Now consider
those $f$'s where $K$ lies between the $0$ and $1$ hyperplanes, and
where $x$ lies in the intersection of $K$ with the $1$ hyperplane.
Then any $y\in \ext{K}$ falls within a unique $t\in [0,1]$ hyperplane.
Finally, consider all such foliations, and take the infimum of the $t$
such that $f(y)=t$.

In some nice cases, there is a unique affine function $f$ such that
$f(x)=1$ and $\min \{ f(y):y\in K\} =0$; thus, $\ratio{K}{x}{y}=f(y)$.
For example, when $K$ is the unit sphere in $\mathbb{R}^{3}$, and
$x,y$ are points on the surface of the sphere, then
$\ratio{K}{x}{y}=\frac{1}{2}(1+x\cdot y)$.  That is, the transition
probability is given by the (normalized) tangent function to $K$ at
$x$.

\begin{figure}[center]
\begin{pspicture}(-6,-2.25)(3,3)
\psset{Beta=5,Alpha=105,linewidth=0.01mm,unit=1.5}
\newrgbcolor{lightblue}{0.68 0.85 0.9}
\pstThreeDSquare[linecolor=lightgray,fillstyle=solid,fillcolor=lightgray](0,0.7071,-0.7071)(+1,0,0)(0,0.7071,0.7071)
\pstThreeDSquare[linecolor=lightgray,fillstyle=solid,fillcolor=lightgray](0,0.7071,-0.7071)(-1,0,0)(0,0.7071,0.7071)
\pstThreeDSquare[linecolor=lightgray,fillstyle=solid,fillcolor=lightgray](0,0.7071,-0.7071)(+1,0,0)(0,-0.7071,-0.7071)
\pstThreeDSquare[linecolor=lightgray,fillstyle=solid,fillcolor=lightgray](0,0.7071,-0.7071)(-1,0,0)(0,-0.7071,-0.7071)
%
\pstThreeDSphere(0,0,0){1.5}
\pstThreeDSquare[linecolor=lightgray,fillstyle=solid,fillcolor=lightgray](0,-0.7071,0.7071)(+1,0,0)(0,0.7071,0.7071)
\pstThreeDSquare[linecolor=lightgray,fillstyle=solid,fillcolor=lightgray](0,-0.7071,0.7071)(-1,0,0)(0,0.7071,0.7071)
\pstThreeDSquare[linecolor=lightgray,fillstyle=solid,fillcolor=lightgray](0,-0.7071,0.7071)(+1,0,0)(0,-0.7071,-0.7071)
\pstThreeDSquare[linecolor=lightgray,fillstyle=solid,fillcolor=lightgray](0,-0.7071,0.7071)(-1,0,0)(0,-0.7071,-0.7071)
\pstThreeDPut(0,-0.7071,0.7071){$\bullet$}
\pstThreeDPut(0,0.7071,-0.7071){$\bullet$}
\pstThreeDPut(0,0.82,-0.82){$x$}
\pstThreeDPut(0,0.7071,0.7071){$\bullet$}
\pstThreeDPut(0,0.82,0.82){$y$}
\pstThreeDPut(0,2.3,-1){$f^{-1}(1)$ hyperplane}
\pstThreeDPut(0,-2.3,1){$f^{-1}(0)$ hyperplane}
\end{pspicture}
\caption{The Bloch sphere}
\label{bloch}
\end{figure}
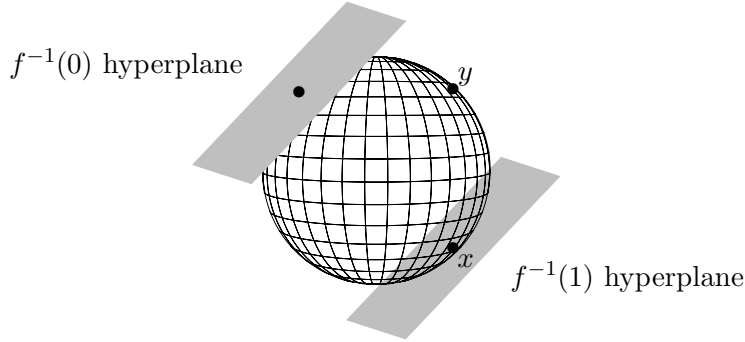

In fact, the sphere is a familiar case from QM: it is affinely
isomorphic to the set of density operators on $\mathbb{C}^{2}$, and
the affine ratio corresponds to the standard quantum-mechanical
transition probability.  Indeed, the equivalence between the two
notions holds quite generally.

\begin{lemma} Let $\hil{H}$ be a complex Hilbert space, and let $K$ be
  the convex set of density operators on $\hil{H}$.  Then for any
  projection operators $E,F\in \ext{K}$, $\ratio{K}{E}{F}=\tr{EF}$.
  \label{transition}
\end{lemma}

\begin{proof} 
  Consider the affine function $f:K\rightarrow [0,1]$ given by
  $f(D)=\tr{ED}$, for all $D\in K$.  We claim that
  $f(F)=\ratio{K}{E}{F}$, for all $F\in \ext{K}$.  For this it will
  suffice to show that for any $g\in A(K)$, if
  $\mathrm{range}(g)\subseteq [0,1]$ and $g(E)=1$, then $g \geq f$.
  Let $g$ be such a function.  Since $A(K)$ is order-isomorphic to
  $\bhs$, there is a self-adjoint operator $A$ on $\hil{H}$ such that
  $A$ has spectrum in $[0,1]$, and $g(F)=\tr{AF}$, for all $F\in
  \ext{K}$.  Let $\ket{\alpha}$ be a unit vector in the range of $E$,
  and let $\ket{\beta}$ be a unit vector in the range of $F$.  Then
  $\langle \alpha |A|\alpha \rangle =g(E)=1$, and it follows that
  $A\ket{\alpha}=\ket{\alpha}$.  By the spectral theorem, $EA=AE=E$,
  and so $\langle \beta |(I-E)A|\beta \rangle \geq 0$.  Therefore,
  \begin{equation} \langle \beta |A|\beta \rangle =\langle \beta
      |EA|\beta \rangle +\langle \beta |(I-E)A|\beta \rangle \geq
      \langle \beta |E|\beta \rangle .\end{equation}
That is, $g(F)\geq f(F)$, for all $F\in \ext{K}$.
\end{proof}

We can now show that the Schr*dinger theory satisfies the no cloning
axiom, but violates the no bit commitment axiom.  First, we claim that
the permissible state transformations of a theory with state space $L$
correspond to affine endomorphisms of $L$.  (A transformation is
reversible iff it is one-to-one.)  If $\eta$ is an affine endomorphism
of $L$, and if $L'=\eta (L)$, then
\begin{equation} \ratio{L'}{\eta (x)}{\eta (y)}\geq \ratio{L}{x}{y}
  ,\qquad \forall x,y\in L .\end{equation}  In the Schr*dinger theory, we have two
subsystems $A,B$ (each qubits) combined in a nonstandard way into a
composite system $AB$.  Suppose for reductio ad absurdum that states
of system $A$ can be cloned by using system $B$ as a cloning machine.  That is, there is a
ready state $x_{0}$ of $B$ and a state transformation $\eta$ on $AB$
such that $\eta (x\otimes x_{0})=x\otimes x$ for all pure states $x$ of
$A$.  Let $x$ and $y$ be non-orthogonal pure states; that is, $0<\ratio{A}{x}{y}<1$.  It then follows that
\begin{equation} \ratio{A}{x}{y}\leq \ratio{AB}{x\otimes x_{0} }{y\otimes
    x_{0} }\leq \ratio{AB}{x\otimes x}{y\otimes
    y}\leq \ratio{A}{x}{y}^{2}, \label{square} \end{equation}  
in contradiction with the assumption $x$ and $y$ are non-orthogonal.  (The first
inequality follows from the fact that $x\mapsto x\otimes
x_{0}$ is an affine embedding of $A$ into $AB$.  The second inequality follows from the fact that the cloning map cannot
decrease transition probabilities.  The final inequality follows
from the fact that $\ratio{AB}{x\otimes x}{y\otimes y}\leq T$, where
    $T$ is the transition probability of $x\otimes x$ and $y\otimes y$
    relative to the full state space of $\twoqubit$; and 
$T=\tp{x\otimes x}{y\otimes y}=|\langle x|y\rangle |^{4}=\ratio{A}{x}{y}^{2}$.) 
Therefore, the cloning map $\eta$ does not exist.
    
In order to see that the Schr*dinger theory allows an unconditionally
secure bit commitment protocol, consider the direct analogue of the
BB84 protocol \cite{bb84}.  In this protocol, Alice encodes bit $0$
into the mixed state $D_{0}=\frac{1}{2}(|01\rangle \langle
01|+|10\rangle \langle 10|)$, and she encodes bit $1$ into the mixed
state $D_{1}=\frac{1}{2}(\braket{\alpha \beta}+\braket{\beta \alpha }
)$, where $\ket{\alpha}=2^{-1/2}(\ket{0}+\ket{1})$ and
$\ket{\beta}=2^{-1/2}(\ket{0}-\ket{1})$.  Since
$\mathrm{Tr}_{A}(D_{0})=\mathrm{Tr}_{A}(D_{1})$, this protocol is
perfectly concealing.  If Alice could prepare the EPR-Bohm state $E$,
i.e., the projection onto the vector $2^{-1/2}(\ket{01}-\ket{10})$,
then this protocol would not be binding; because $E$ can be
transformed by local nonselective operations into either $D_{0}$ or
$D_{1}$.  However, in the Schr*dinger theory, there are no such
entangled states; indeed, there is no state that Alice can transform
into either $D_{0}$ or $D_{1}$.  Therefore, this protocol is perfectly
binding.

Since the Schr*dinger theory prohibits cloning, but allows
unconditionally secure bit commitment, it follows from
Theorem~\ref{independence} that it does not admit a $C^{*}$-algebraic
formulation.  We devote the next section to establishing a stronger
claim: the Schr*dinger theory does not admit a JB algebraic
formulation.

\subsection{Pathology of the Schr*dinger theory}

By using the generalized definition of transition probability for
arbitrary convex sets, we can see classical transition probabilities
are always in $\{ 0,1\}$, whereas quantum mechanical transition
probabilities can lie anywhere in the unit interval.  We will use the
transition probability to define a generalized notion of the
superposition of two pure states.

\begin{defn} Pure states $x,y\in \ext{K}$ are said to be {\it orthogonal} just in case
  $\ratio{K}{x}{y}=0$.  Two orthogonal states $x,y\in \ext{K}$ are
  said to be \emph{superposable in} $K$ just in case there is a $z\in
  \ext{K}$ such that $\ratio{K}{x}{z}=\frac{1}{2}=\ratio{K}{y}{z}$.
\end{defn}

This definition is motivated by the following considerations.  If
observables correspond to affine functions on $K$ (as is usually
assumed in the convex sets approach), a measurement designed to
distinguish $x$ from $y$ can be represented by an affine function
$f:K\rightarrow [0,1]$, where $f(x)=1$ and $f(y)=0$.  If there is such
a function $f$, then $\ratio{K}{x}{z}=\frac{1}{2}=\ratio{K}{y}{z}$ iff
$f(z)=\frac{1}{2}$.  That is, when the system is in state $z$, the $x$
and $y$ outcomes of an $f$-measurement are equally likely.

For the case of JB algebra state spaces, if two pure states can be
connected by a continuous path, then they can be coherently
superposed.\footnote{According to \citeA{hardy}, QM is differentiated
  from classical probability theory by the assumption that there is a
  continuous transition between any two pure states.  Hardy's claim is
  true in the JB algebraic framework (if we take the relevant topology
  to be the norm topology), if we think of ``quantum'' systems as
  those systems that have a single non-trivial superselection sector,
  and ``classical'' systems as those systems whose superselection
  sectors are singletons.}

\begin{lemma} 
  Let $K$ be the state space of a JB algebra, and let $x,y$ be
  orthogonal states in $\ext{K}$.  If $x$ and $y$ are connected by a
  norm-continuous path in $\ext{K}$, then $x$ and $y$ are superposable
  in $K$.
  \label{super}
\end{lemma}

\begin{proof} 
  Suppose that $x,y\in \ext{K}$ are orthogonal, and that $x$ and $y$
  are connected by a norm-continuous path in $\ext{K}$.  Let
  $F=\face{x}{y}$.  By part~\ref{path} of the Root Theorem, there is
  an affine isomorphism $\phi$ from $F$ onto $B^{n}$, with $n\geq 2$.
  Let $\{ e_{1},e_{2},\dots ,e_{n}\}$ be the canonical orthonormal
  basis for $\mathbb{R}^{n}$.  Since there is an affine automorphism
  of $B^{n}$ that maps $\phi (x)$ to $e_{1}$, we may suppose that
  $\phi (x)=e_{1}$.  An exercise in elementary geometry shows that
  $\phi (y)=-e_{1}$.  ($\phi$ preserves affine ratios, and $-e_{1}$ is
  the unique $r\in B^{n}$ such that $\ratio{B^{n}}{e_{1}}{r}=0$.)
  Furthermore,
  $\ratio{B^{n}}{e_{1}}{e_{2}}=\ratio{B^{n}}{-e_{1}}{e_{2}}=\frac{1}{2}$.
  Thus, if we choose $z=\phi ^{-1}(e_{2})$, then $\ratio{K}{x}{z}\geq
  \ratio{F}{x}{z}=\frac{1}{2}$ and $\ratio{K}{y}{z}\geq
  \ratio{F}{y}{z}=\frac{1}{2}$.
\end{proof}

The Schr*dinger theory seems to have some non-trivial superselection
rule, because it does not seem to allow coherent superpositions of,
say, $\ket{01}$ and $\ket{10}$.  We begin by confirming that such
states are not superposable.

\begin{lemma} Let $K$ be the set of separable states of
  \mbox{$\twoqubit$}.  Then $\ket{01}$ and $\ket{10}$ are not
  superposable in $K$.
  \label{specific} \end{lemma}

\begin{proof} 
  Let $x=\braket{01}$, let $y=\braket{10}$, and suppose for reductio
  ad absurdum that there is a $z=\ket{\alpha \beta}=\ket{\alpha
  }\ket{\beta }\in \ext{K}$ such that
  $\ratio{K}{x}{z}=\frac{1}{2}=\ratio{K}{y}{z}$.  If $L$ is the full
  state space of $\twoqubit$, then $\ratio{K}{v}{w}\leq
  \ratio{L}{v}{w}$ for any $v,w\in \ext{K}$ (since $\ext{K}\subseteq
  \ext{L}$).  Thus, Lemma~\ref{transition} entails that
\begin{eqnarray} 1&=&\ratio{K}{x}{z}+\ratio{K}{y}{z} \\
&\leq & \ratio{L}{x}{z}+\ratio{L}{y}{z} \\
&=& |\langle 01|\alpha \beta \rangle |^{2}+|\langle 10|\alpha \beta \rangle |^{2}\: \leq \:
  1 .\label{burb} \end{eqnarray}
We now show that either $\tp{01}{\alpha \beta }=0$ or
$\tp{10}{\alpha \beta }=0$.  For this, let
  \begin{equation} a=\tp{0}{\alpha},\quad b=\tp{1}{\beta},\quad
    c=\tp{0}{\beta},\quad d=\tp{1}{\alpha} .\end{equation}
  Thus, Eqn.~\ref{burb} becomes $ab+cd=1$.  Since $\{
  \ket{0},\ket{1}\}$ is an orthonormal basis for $\mathbb{C}^{2}$, we
  also have $b=1-c$ and $d=1-a$.  Hence, $a+c-2ac=1$.  The functions
  $[0,1]\ni a\mapsto a+c-2ac$ (for fixed $c\in [0,1]$) and $[0,1]\ni
  c\mapsto a+c-2ac$ (for fixed $a\in [0,1]$) are affine.  Thus,
  $a+c-2ac$ achieves its maximum value only at extreme points of the
  convex set $[0,1]\times [0,1]$.  Checking these points, we find that
  $a+c-2ac\leq 1$, with equality achieved only when $(a,c)=(1,0)$ or
  $(a,c)=(0,1)$.  If $c=0$, then $\tp{10}{\alpha \beta }=cd=0$.
  Similarly, if $a=0$, then $\tp{01}{\alpha \beta }=ab=0$.  Applying
  Lemma~1 again, it follows that either
\begin{equation}
\ratio{K}{x}{z} \leq \ratio{L}{x}{z} =\tp{01}{\alpha \beta}=0 ,\end{equation} or
\begin{equation}
\ratio{K}{y}{z} \leq \ratio{L}{y}{z} =\tp{10}{\alpha \beta}=0
,\end{equation}  
both of which
  contradict our assumption that $z$ is an equally weighted 
superposition of $x$ and $y$.  Therefore, $\ket{01}$ and $\ket{10}$
are not superposable in~$K$.  \end{proof}

The previous Lemma shows that if the Schr*dinger theory permits a JB
algebraic formulation, then the states $\ket{01}$ and $\ket{10}$ must
lie in different superselection sectors.  However, $\ket{01}$ and
$\ket{10}$ are connected by a continuous path of pure product
states.\footnote{Since $K$ is not a topological space, this statement
  doesn't really make sense.  However, since $K$ is a subset of
  $\mathbb{R}^{n}, (n<\infty )$, there is a unique topology $\tau$ on
  $K$ that is compatible with its affine structure; namely, the
  relative topology from $\mathbb{R}^{n}$.  So, if $K$ were isomorphic
  to a JB algebra state space, the transported norm topology would be
  equivalent to $\tau$.}

\begin{lemma} Let $K$ be the set of separable states of
  \mbox{$\twoqubit$}.  Then there is a continuous path in $\ext{K}$
  between $\ket{01}$ and $\ket{10}$.  \label{connected}
\end{lemma}

\begin{proof} 
  By symmetry, and since path-connectedness of points is transitive,
  it will suffice to show that there is a norm-continuous path in
  $\ext{K}$ between $\ket{01}$ and $\ket{11}$.  Let $\hsnorm{\cdot}$
  denote the Hilbert-Schmidt norm on $\nbit$; i.e.,
  $\hsnorm{A}=\tr{A^{*}A}^{1/2}$.  There is a
  $\hsnorm{\cdot}$-continuous function $f$ from $[0,1]$ into the set
  of one-dimensional projections on $\mathbb{C}^{2}$ such that
  $f(0)=\braket{0}$ and $f(1)=\braket{1}$.  Define a function
  $g:[0,1]\rightarrow \ext{K}$ by setting $g(t)=f(t)\otimes
  \braket{1}$.  Since $\hsnorm{A\otimes B}=\hsnorm{A}\hsnorm{B}$ for
  all operators $A,B$ on $\mathbb{C}^{2}$, it follows that
  \begin{equation} \hsnorm{g(t)-g(t')} =
  \hsnorm{(f(t)-f(t'))\otimes \braket{1} \, } 
=\hsnorm{f(t)-f(t')} ,\end{equation} for all $t,t'\in
  [0,1]$.  Therefore, $g$ is $\hsnorm{\cdot}$-continuous as a mapping into
  $\ext{K}$ with the relative topology inherited from the state space
  of $\twoqubit$.
\end{proof}

The previous two Lemmas show that the topology and affine structure of
the separable state space do not mesh in the way that these structures
mesh in JB algebra state spaces.

\begin{thm} 
  The set of separable states of $\twoqubit$ is not affinely
  isomorphic to the state space of a JB algebra.
  \label{schro}
\end{thm}

\begin{proof} 
  Suppose for reductio ad absurdum that $K$ is affinely isomorphic to
  the state space of a JB algebra.  Let $x=\braket{10}$ and let
  $y=\braket{01}$.  By Lemma~\ref{connected}, there is a continuous
  path between $x$ and $y$, and so Lemma~\ref{super} entails that $x$
  and $y$ are superposable in $K$.  But this contradicts
  Lemma~\ref{specific}.  Therefore, $K$ is not affinely isomorphic to
  the state space of a JB algebra.
\end{proof}

This result shows that the simplest Schr{\"o}dinger-like theory ---
viz., the Schr*dinger theory --- does not admit a JB algebraic
formulation.  Thus, it provides some evidence for the claim that even
within the JB algebraic framework, locally quantum mechanical theories
have nonlocally entangled states; and it suggests that even within the
JB algebraic framework, the no cloning axiom entails the no bit
commitment axiom.

The upshot, then, of this section is to confirm worries that the
$C^{*}$ assumption is doing too much work in the CBH theorem; and,
furthermore, it probably wouldn't help matters if we were to derive a
generalized CBH theorem for JB algebras.  But, of course, there is
still hope that within a suitably broader mathematical framework
(e.g., Segal-algebras (see \citeNP{segal}), and the dual theory of
spectral convex sets), the three axioms are independent, and together
entail QM. 

\section{Conclusion}

This note attempts to clarify the limits of recent
information-theoretic characterizations of QM.
However, in doing so, it has raised a number of further questions,
both of a technical and a philosophical nature.  

First, we conjecture that the three information-theoretic axioms are
independent in the Segal-algebraic framework, and that the conjunction
of the axioms entails QM.  This generalized version of the CBH theorem
would not only address the worries raised in the previous section, but
might also help shed light on traditional questions, such as physical
reasons for using complex coefficients rather than reals or
quaternions.
 
Second, the considerations in this paper suggest that we take a closer
look at different ways of putting together composite systems, where
all systems are assumed to have convex state spaces.  It is known that
there are several different notions of the ``tensor product'' of
compact convex sets (see, e.g., \citeNP{namioka}).  Thus, it would be
interesting to see which of these products preserve which
information-theoretic properties of the component systems.  More
specifically, suppose that $\otimes$ is a tensor product of compact
convex sets that preserves the defining properties of JB algebra state
spaces.  Then does it follow that $K\otimes L$ has nonlocally
entangled states whenever $K$ and $L$ are not simplexes?  Or does the
JB algebraic framework permit the existence of a Schr{\"o}dinger-like
theory?  If the JB algebraic framework does not, does the broader
Segal-algebraic framework permit the existence of a
Schr{\"o}dinger-like theory?

Finally, our discussion has raised the question of the role of
constraints (either {\it a priori} or operational) on theory
construction.  On the one hand, if there are no constraints on theory
construction --- i.e., if there is no minimum amount of mathematical
structure shared by all theories, and if any fairy tale can count as a
legitimate ``toy theory'' --- then it would be hopeless to try to {\it
  derive} QM from information theoretic principles, or
from any other sort of principles for that matter.  (E.g., why assume
that the results of measurements are real numbers?  Why assume that
measurements have single outcomes?  Why assume that the laws of
physics are the same from one moment to the next?)  On the other hand,
the idea that it is legitimate to assume a fixed background framework
for physical theories seems to come into tension with the empiricist
attitude that drove the two major revolutions in physics in the 20th
century; and the last thing we want is to impede the search for a
future theory that would generalize QM. 

\newpage
\begin{center} {\bf Acknowledgments} \end{center}

Many of the ideas in this paper originated from conversations with
Jeff Bub.  Thanks also to Rob Spekkens and to an anonymous referee for
comments on an earlier draft.

\nocite{hardy}
\bibliographystyle{apacite}
\bibliography{bib}

\end{document}